\documentclass[prb,aps,showpacs,unsortedaddress,superscriptaddress,twocolumn]{revtex4}
\usepackage{graphicx}

\begin{document}

\title{Quantum vortex dynamics in two-dimensional neutral superfluids}

\author{C.-C. Joseph Wang}
\email{joseph@physics.utexas.edu}
\homepage{http://www.ph.utexas.edu/~joseph/joseph.html}

\affiliation{The University of Texas at Austin, Department of
Physics, 1 University Station C1600, Austin, TX 78712-0264}

\author{R.A. Duine}
\email{R.A.Duine@uu.nl} \homepage{http://www.phys.uu.nl/~duine}

\affiliation{Institute for Theoretical Physics, Utrecht
University, Leuvenlaan 4, 3584 CE Utrecht, The Netherlands}

\author{A.H. MacDonald}
\email{macd@physics.utexas.edu}
\homepage{http://www.ph.utexas.edu/~macdgrp}

\affiliation{The University of Texas at Austin, Department of
Physics, 1 University Station C1600, Austin, TX 78712-0264}

\date{\today}

\begin{abstract}
We derive an effective action for the vortex position
degree-of-freedom in a superfluid by integrating out condensate
phase and density fluctuation environmental
modes. When the quantum dynamics of environmental fluctuations is
neglected, we confirm the occurrence of the vortex Magnus force
and obtain an expression for the vortex mass. We find that this
adiabatic approximation is valid only when the superfluid droplet
radius $R$, or the typical distance between vortices, is very much
larger than the coherence length $\xi$. We go beyond the adiabatic
approximation numerically, accounting for the quantum dynamics of
environmental modes and capturing their dissipative coupling to
condensate dynamics.  For the case of an
optical-lattice superfluid we demonstrate that vortex motion
damping can be adjusted by tuning the ratio between the tunneling
energy $J$ and the on-site interaction energy $U$.  We comment on
the possibility of realizing  vortex-Landau-level physics.
\end{abstract}

\maketitle

\def\bx{{\bf x}}
\def\bk{{\bf k}}
\def\br{{\bf r}}
\def\bu{{\bf u}}
\def\half{\frac{1}{2}}
\def\args{(\bx,t)}

\section{Introduction}
Superfluids\index{superfluid} accommodate angular momentum
by nucleating quantized vortices.\cite{feynman1955,onsager1949}
The rapid progress of research on Bose-Einstein condensed
ultracold atoms in magnetic traps\cite{anderson1995,Ketterle1995,Leggett,Dalfovo} and optical
lattices\cite{Greiner,RMP} over the past decade has opened up an opportunity to study vortex
physics in a new setting.\cite{madison2000,raman2001,haljan2001,hodby2002}
New possibilities for exploring phenomena in which the quantum behavior of
vortices is important are particularly attractive.  It is known, for
instance, that quantum fluctuations induce
vortex lattice\index{vortex lattice melting} melting\cite{VLatMelt} when the number of atoms per vortex is less than
around ten, and that Boson quantum Hall states\cite{FQH}
(in which vortices form incompressible quantum fluids) emerge at still lower atom densities
(for a recent review, consult Ref.~[\onlinecite{Cooper}]).
Phase separation between different vortex states driven by
inhomogeneity in the condensate density\cite{Fischer}
and vortex-Peierls states\index{Vortex-Peierls states} in some optical lattices are
manifestations of quantumness in vortex physics that is enhanced by geometrical
frustration.\cite{A. Burkov}

In spite of the long history of vortex physics in neutral
superfluids there are still some controversies surrounding the
understanding of vortex quantum dynamics.\cite{Ao2005}  The main
concerns center on the sources of forces\index{forces} on a
quantum vortex and on the effective vortex mass in various
circumstances\index{vortex mass in various
circumstances}.\cite{Kopnin,Ao1999,Niu1994,Fischer2} The consensus
view(for completeness, however, see also Ref.~[\onlinecite{E. B. Sonin}]),
based on either Berry phase\index{Berry phase}
arguments\cite{Thouless 1993} or hydrodynamic
theory\cite{hydrodynamic}, is that a vortex has a finite mass and
experiences a Magnus force\index{Magnus force} proportional to the
boson density in its vicinity. According to these theories, then,
a vortex behaves like a massive charged particle in an effective
magnetic field due to the background
bosons.\cite{Simanek,fisher_duality} Vortex kinetic energy should
therefore be subject to the same Landau quantization conditions
which qualitatively alter the properties of electrons subject to
an external magnetic field. Vortex Landau quantization has
nevertheless never been observed.  A key aim of this work is to
help point the way toward circumstances in which this distinctive
aspect of vortex quantum mechanics is more likely to be exhibited.

In early work on the properties of vortices, the vortex mass $m_v$ was often
treated as a purely phenomenological parameter.\cite{DONNELLY}
More attention has been given to the microscopic underpinning of
vortex mass in recent work.
For example, Duan\cite{Duan1993} demonstrated that the vortex mass
is generated by the superfluid compressibility, although the
Magnus force was not considered. Arovas and Freire
\cite{arovas1997} formulated a theory of vortex mass by assuming that the
coupling between the vortex position and environment is dictated
by gauge invariance as in electromagnetism, with the vortex
playing the role of the charged particle, and phonons that of the
electromagnetic field. In their approach the resulting minimal
coupling leads to phonon emission by a moving vortex, which gives
the vortex a frequency-dependent effective mass. Further progress
in clarifying the microscopic physics of the vortex mass was
achieved more recently in Ref.~[\onlinecite{Thouless PRL}], which
approached the issue by using perturbation theory with a rotating
pinning potential. These authors confirmed that the vortex mass is
due to density fluctuations in the superfluid and showed that the
value of the mass can depend on details of the vortex core
structure.

In this paper, we use a different approach to derive
an effective action for an isolated vortex.
We start with a careful identification of the
vortex-translation zero-modes in the Gaussian fluctuation
spectrum of a superfluid which contains a vortex.
We then integrate out all the remaining Gaussian fluctuations,
which consist of superfluid density and phase fluctuations distorted by the presence of the vortex, to obtain an
effective theory which depends only on the vortex position degree-of-freedom.
In the adiabatic limit, in which the quantum dynamics of the
environmental fluctuations can be ignored, we reproduce the vortex mass predicted by
Duan.\cite{Duan1993} We confirm that the appearance of the vortex
mass is due primarily to the coupling between vortices and
condensate density fluctuations outside the vortex core.
When quantum fluctuations of the environment modes are
allowed, we find that the picture in which the vortex exists as a
well-defined quantum particle with an energy-independent mass
holds only when the vortices are in a
in superfluid cloud with radius $R$ ,
very much larger than the coherence length $\xi$.
In a system with many vortices the corresponding condition requires that the
typical distance between vortices be very much larger than $\xi$.

We consider, in addition, the situation of a vortex in an optical
lattice condensate and calculate the vortex mass and effective
magnetic field numerically.
Numerical results for the
vortex mass tend to agree reasonably well with analytical results based on a continuum
model.

Numerical results for the effective magnetic field, however,
depend strongly on the optical lattice and do
not compare favorably with analytical results.
For large vortex core sizes the discrepancy may be explained as a
consequence of the large fraction of the system area that
is occupied by vortex cores.  For small vortex cores, on the other hand,
lattice-pinning by the vortex core plays an important role.
Taken together, these two effects
imply that sharp vortex-cyclotron resonances and vortex-Landau-level
physics are difficult to observe in optical-lattice Bose condensates.

The remainder of this paper is organized as follows. In
Sec.\ref{sec:theory} we outline our model and present our approach
to the evaluation of the vortex mass and effective magnetic field. We
develop this approach analytically for the continuum limit. In
Sec.~\ref{sec:OL} we present results of numerical calculations for
lattice bosons. We end with conclusions in
Sec.~\ref{sec:conclusion}.

\section{Theoretical formulation}
\label{sec:theory}
\subsection{Effective action}
\index{effective action}
We consider a dilute interacting bosonic gas in a reduced
two-dimensional geometry. Our starting point is the partition
function\index{partition function} for the two-dimensional dilute
gas with contact interaction $g>0$, written as a coherent-state
path integral over the fields $\phi^* ({\bf x}, \tau)$ and $\phi
({\bf x}, \tau)$. It is given by
\begin{equation}
\label{eq:partfunc1}
  Z = \int d[\phi^*] d[\phi] \;  e^{-S[\phi^*,\phi]/\hbar}~,
\end{equation}
with the Euclidean action $S[\phi^*,\phi]=\int_{0}^{\hbar \beta}
d\tau L$ and Lagrangian
\begin{widetext}
\begin{equation}
\label{eq:action1}
  L[\phi^*,\phi] = \int d{\bf x} \;
  \phi^* ({\bf x}, \tau) \left[
    \hbar \frac{\partial}{\partial \tau}
    - \frac{\hbar^2 \nabla^2}{2 m} - \mu + \frac{g|\phi({\bf x},\tau)|^2}{2} \right] \phi{({\bf x},\tau) } \; = \; \int d{\bf x} \; \left[
     \phi^* ({\bf x}, \tau) \hbar \frac{\partial}{\partial \tau}  \phi{({\bf x},\tau)}  + {\cal E}[\phi({\bf x})]
     \right]~.
\end{equation}
\end{widetext}
The functional integration in Eq.~(\ref{eq:partfunc1}) is over all fields
periodic on the imaginary-time axis between $\tau=0$ and
$\tau=\hbar \beta \equiv \hbar/k_B T$, where $k_BT$ is the thermal
energy.  The integral in Eq.~(\ref{eq:action1}) is over
two-dimensional position {\bf x}. In the context of cold atoms,
the model can be realized with a sufficiently strong harmonic
trapping potential with frequency $\omega_z$ in the $z$-direction.
The effective contact interaction strength $g$ is related to the
$s$-wave scattering length $a$\index{$s$-wave scattering length}
of the atoms by $g=4 \pi a \hbar^2/(m d_z)$, with $m$ the mass of
an atom. The extent of the system in the $z$-direction is denoted
by $d_z$ and is related to the strong harmonic trapping potential
in that direction given by $d_z \sim \sqrt{\hbar/m\omega_z}$. We
ignore the effects of the radial trapping potentials for now in
order to avoid unessential complications.

We consider field configurations close to that of a classical
vortex with counterclockwise phase winding $+2\pi$ located at
${\bf r}_v$:
\begin{equation}
\label{eq:separation}
  \phi ({\bf x}, \tau) = \phi_v ({\bf x} - {\bf r}_v (\tau)) + \delta \phi
  ({\bf x},\tau).
\end{equation}
Here, $\phi_v({\bf x}-{\bf r}_v)$ is a local minimum of the energy
functional ${\cal E}[\phi({\bf x})]$ in Eq.~(\ref{eq:action1}{)
for any value of the vortex position ${\bf r}_v$. Equivalently we
can say that $\phi_v ({\bf x})$ is the stationary solution of the
Euler-Lagrange equation\index{Euler-Lagrange equation} of the
action in Eq.~(\ref{eq:action1}), ({\em i.e.}, $\delta S
[\phi^*_v,\phi_v]/\delta \phi^*=0$) containing one vortex at the
origin. Vortex translation in time is described by ${\bf r}_v
(\tau)$. However, the fluctuating part of the field $\delta \phi
({{\bf x}} ,\tau)$ contains zero modes\index{zero modes} that also
describe the translation of the vortex. For example, a fluctuation
$\delta \phi \propto
\partial_x \phi_v$ describes translation of the vortex in the
$x$-direction. Generally, we denote the zero modes responsible for
translation of the vortex in the $x$ and $y$-direction by
$\phi_{0i}$, with $i \in \{ x,y\}$. To avoid double-counting in the
functional integral we have to enforce the constraint that the fluctuations
$(\delta \phi^*,\delta \phi)$ are orthogonal to the zero modes.
Hence, the fluctuations are required to obey

\begin{equation}
\label{eq:contraint}
  \langle \phi_{0i}| \delta \phi \rangle \equiv
  \int d{\bf x} \; \phi^*_{0i} ({{\bf x}}) \delta \phi ({{\bf x}}, \tau)
   = 0~.
\end{equation}
This procedure precisely separates vortex-position and what
we will refer to as environmental fluctuations.
To enforce the above constraint and, at the same time, introduce
the vortex position as a dynamical variable in the function
integral we use the Fadeev-Popov procedure\index{Fadeev-Popov
procedure} \cite{gervais1975,braun1996}. We rewrite the partition
function as
\begin{equation}
\label{eq:partfunc2}
  Z = \int d[{\bf r}_v] d [\phi^*] d [\phi]
  \delta \left( \langle \phi_{0x}| \delta \phi \rangle \right)
  \delta \left( \langle \phi_{0y}| \delta \phi \rangle \right)
  e^{-S[{\bf r}_v, \delta\phi^*,\delta\phi]/\hbar}~,
\end{equation}
and expand the action up to quadratic order in the fluctuations.
The resulting actions consists of three parts, a bare vortex motion contribution
$S_v$, an environmental-fluctuation contribution $S_e$, and a
coupling term $S_c$:
\begin{equation}
  S[{\bf r}_v, \delta\phi^*,\delta\phi]
  = S_v[{\bf r}_v] + S_e [\delta\phi^*,\delta\phi] + S_c[{\bf r}_v,
  \delta\phi^*,\delta\phi]~.
\end{equation}

Because ${\cal E}[\phi_v({\bf x}-{\bf r}_v)]$ is independent of
${\bf r}_v$, the bare vortex contribution to the action is due
entirely to the kinetic terms:
\begin{widetext}
\begin{equation}
\label{action:v}
 S_v [{\bf r}_v]=\langle \phi_{\nu}|\hbar\partial_{\tau}\phi_{\nu}\rangle
 = \frac{1}{2}[\langle \phi_{\nu}|\hbar\partial_{\tau}\phi_{\nu}\rangle-
 \langle \hbar \partial_{\tau}\phi_{\nu}|\phi_{\nu}\rangle]
=i \hbar \Big( \int d{\bf x} \, {\rm Im} \left[\frac{\partial
\phi_{\nu}^{*}({\bf x})}{\partial x} \frac{\partial
\phi_{\nu}({\bf x})}{\partial y}\right] \Big) \int d\tau [x_{\nu}{\dot
y_{\nu}}-y_{\nu}{\dot x_{\nu}}].
\end{equation}
The spatial integral factor in Eq.~(\ref{action:v}), is readily
evaluated for a circularly-symmetric vortex wave function
$\phi_{v}({\bf x}) = f(r) \exp(i\theta)$ using polar coordinates:
\begin{equation}
\label{eq:Berryxy} \int d{\bf x} {\rm Im}\left[\frac{\partial
\phi_{\nu}^{*}({\bf x})}{\partial x} \frac{\partial
\phi_{\nu}({\bf x})}{\partial y}\right] = 2 \pi \int dr f
\frac{df}{dr} = \pi f^2(\infty) = \pi n_c~,
\end{equation}
\end{widetext}
where $n_c = |\phi_v|^2$ is the condensate density in the uniform superfluid far from the vortex.
As a result, we obtain
\begin{equation}
\label{eq:s0}
  S_v [{\bf r}_v] = i \hbar \pi n_c \int_0^{\hbar\beta} d \tau \left[
     \epsilon^{ij} r_v^i (\tau) \frac{d r_v^j (\tau)}{d \tau}
   \right]~,
\end{equation}
where $\epsilon^{ij}$ is the two-dimensional Levi-Civita tensor
and summation over repeated Cartesian indices $i,j \in \{x,y\}$ is
implied.  After a Wick rotation from imaginary to real time ($\tau
\longrightarrow it$), this contribution to the action can be
identified with the magnetic potential which appears for charged
particles in a magnetic field.  When added to a free particle
Lagrangian it implies kinetic energy quantization and Landau level
formation.  The effective magnetic length $\ell \equiv (\hbar
c/qB)^{1/2}$ corresponding to this effective magnetic potential is
$\ell_{\rm eff} = (2 \pi n_{c})^{-1/2}$.  The effective
Aharonov-Bohm phase accumulated by a vortex wave function
surrounding an area $A$ is therefore $A/\ell^2 = 2 \pi N_{c}$
where $N_{c}= A n_{c}$ is the number of enclosed condensate
particles.  The correspondence between this phase and the phase $2
\pi N_v$ experienced by bosons enclosing $N_v$ vortices is the
origin of the boson-vortex duality properties which are
useful\cite{Simanek,fisher_duality} in analyzing
superconductor-insulator phase transitions.

It is convenient to express the quadratic action for quantum fluctuations\index{quantum
fluctuations} of the environment $S_e [\delta\phi^*,\delta\phi]$ in the form
\begin{widetext}
\begin{eqnarray}
\label{eq:actionsb}
 &&S_e= \frac{1}{2} \int_0^{\hbar \beta} d \tau \int d {\bf x}
  \; \delta \Psi^{\dag}({\bf x})
  \cdot\!
  \left[\left(
    \begin{array}{ccc}
      \hbar \frac{\partial}{\partial \tau} && 0 \\
       0 &&  -\hbar \frac{\partial}{\partial \tau}
    \end{array}
   \right) +{\mathcal H}\right]
 \!\cdot\! \;
  \delta\Psi({\bf x})
  ,  \nonumber \\
 && ~
\end{eqnarray}
where the vector $\delta \Psi \equiv \left[
      \delta\phi ({\bf x},\tau),
      \delta\phi^* ({\bf x},\tau)
\right]^T$ and the hermitian Hamiltonian ${\mathcal H}$
\begin{eqnarray}
\label{eq:hamflucs}
 {\mathcal H} = \left(
    \begin{array}{ccc}
      - \frac{\hbar^2 \nabla^2}{2 m} - \mu + 2 g|\phi_v({\bf x})|^2 && g \left( \phi_v ({\bf x})\right)^2 \\
       g \left( \phi_v^* ({\bf x})\right)^2  && - \frac{\hbar^2 \nabla^2}{2 m} - \mu + 2 g|\phi_v({\bf x})|^2
    \end{array}
   \right)~.
\end{eqnarray}
The action contribution that describes coupling between vortex translation and
and the quantum fluctuations\index{coupling between vortex zero modes
and fluctuations} of its condensate environment is
\begin{equation}
S_c=\frac{\hbar}{2} \int_0^{\hbar
\beta} d\tau d {\bf x} \; [\partial_i \phi^*_{v} ({\bf x}) \delta
\phi({\bf x},\tau)  - \delta \phi^*({\bf x}, \tau)
\partial_i \phi_{v} ({\bf x})] \;  \dot{r}_v^{i}(\tau)~.
\end{equation}
\end{widetext}
Only kinetic Berry-phase coupling terms appear here because the vortex-translation mode is
a (zero-energy) eigenfunction of the hermitian Hamiltonian ${\mathcal H}$.

In order to focus on vortex properties we next integrate out the environmental fluctuations $\delta \phi^*$
and $\delta \phi$. The constraints imposed by the delta-functional in the
functional integral (Eq.~(\ref{eq:partfunc2})) are conveniently applied by expanding the
fluctuation fields in the basis of eigenfunctions of the Hamiltonian ${\mathcal H}$:
\begin{equation}
\bf{\chi}_\alpha \equiv
 \left(
   \begin{array}{c}
    \chi^1_\alpha  \\
    \chi^2_\alpha   \\
   \end{array}
 \right)
\end{equation}
We expect $\cal{H}$ to have three zero-energy eigenfunctions. The
first ${\bf \chi}_{0p} = (\phi_v, -\phi^*_v)^{T}$ is the global
phase mode\index{global phase mode} that results from the $U(1)$
invariance of the action in Eq.~(\ref{eq:action1}). This mode
makes no contribution to the dynamics of the vortex as will be
shown later. The other two easily identifiable  modes on which we
focus are strictly speaking zero-modes only in the absence of
vortex pinning, and represent vortex translation in orthogonal
directions in the $\hat{x}-\hat{y}$ plane. Hereafter, we will
refer to these translational modes as zero modes in the
discussion. Suitable linear combinations correspond to
translations along translation along any direction in the $x-y$
plane. We choose as basis functions ${\bf \chi}_{0i\in\{x,y\}}
\propto (\partial_i \phi_v,
\partial_i \phi_v^*)^{T}$ where $i$=$x,y$.  One can demonstrate
by direct substitution that these are indeed zero-energy
eigenfunctions of ${\mathcal H}$.  All other fluctuations can be
expanded in terms of the finite-energy $\epsilon_{\alpha}$
eigenfunctions of $\cal{H}$ and correspond to density and phase
fluctuations of a superfluid which has been distorted by the
presence of a vortex.  For the purpose of later discussions, we
define the quantity $\Gamma_{\alpha'\alpha}$, an inner product
between eigenfunctions $\alpha',\alpha$, as
$\Gamma_{\alpha'\alpha}=\frac{1}{2}\langle\chi_{\alpha'}^{\dagger}|\sigma_{z}|\chi_{\alpha}\rangle$,
where $\sigma_{z}$ is the $z$-th Pauli matrix, with the hermitian
property $\Gamma_{\alpha'\alpha}=\Gamma_{\alpha\alpha'}^{*}$. For
example, the Berry-phase coupling between the $x$ and $y$
components of the vortex displacement, considered in
Eq.~(\ref{eq:Berryxy}), is $\Gamma_{xy}$ which is purely
imaginary.

\subsection{Adiabatic vortex dynamics and effective mass}
Now we consider vortex dynamics in the adiabatic
approximation\index{adiabatic approximation}. This amounts to
neglecting the explicit time dependence of environmental modes in
the action of fluctuations $S_{e}[\delta\phi^{*},\delta\phi]$. We
expand the microscopic action $S$ in the basis of fluctuation
Hamiltonian ${\mathcal H}$ eigenfunctions; that is,  $\left[
                              \delta {\bf \phi}({\bf x},\tau),
                              \delta {\bf \phi}^{*}({\bf x},\tau)\\
                          \right]^T
= \sum_\alpha c_\alpha (\tau) {\chi}_\alpha ({\bf x})$ where the
condition $\int d {\bf x} {\chi}_\alpha ({\bf x})^{\dag}{\chi}_{\alpha'}
({\bf x})=\delta_{\alpha,\alpha'}$ is satisfied. In the adiabatic approximation
\begin{equation}
 S_e [c^*,c] = \frac{1}{2}\int_0^{\hbar \beta} d\tau\sum_\alpha
 \epsilon_\alpha \; c^*_\alpha (\tau) c_\alpha (\tau)~.
\end{equation}
The part of the action that describes the coupling becomes
\begin{equation} S_c = \frac{\hbar}{2} \int_0^{\hbar \beta} d\tau
\left[ \sum_\alpha  \Gamma_{i
\alpha}^* c_\alpha (\tau)-c^*_\alpha (\tau) \Gamma_{i\alpha}  \right] \dot{
r}_v^i(\tau)~,
\end{equation}
with $\Gamma_{i\alpha} \equiv \frac{1}{2}\int d{\bf x} \left[ \chi^{1*}_\alpha ({\bf x})
  \partial_i \phi_{v} ({\bf x}) -  \chi^{2*}_\alpha ({\bf x})
  \partial_i \phi_{v}^{*}({\bf x}) \right]~.$  Note that the above expression
  is written in terms of $c_{\alpha}$ and $c^{*}_{\alpha}$ by using the relation that $\delta \phi^{*}({\bf x},\tau)$ is the complex conjugate
  of $\delta \phi({\bf x},\tau)$.

Since the fluctuation action is Gaussian, we can perform the integral over the
environment fields
$c_\alpha^* (\tau)$ and $c_\alpha (\tau)$ analytically.
The effective action of the
vortex is then given by
\begin{equation}
\label{eq:seff}
  S_{\rm eff} [{\bf r}_v] = \int_0^{\hbar\beta} d \tau \left[
   2 \pi i n_c \hbar \frac{1}{2}\epsilon^{ij} r_v^i (\tau) \dot{r}_v^j (\tau)
   +\frac{1}{2} m_v {{\dot{r}_v^{i}}} {{\dot{r}_v^{i}}}\right]~,
\end{equation}
with the effective mass of the vortex\index{effective mass of the vortex} given by
\begin{equation}
\label{eq:massadiabatic}
 m_v=  \hbar^2 \sum_\alpha' \frac{\left|
 \Gamma_{x\alpha}\right|^2}{\epsilon_{\alpha}}= \hbar^2 \sum_\alpha'
 \frac{\left|
 \Gamma_{y\alpha}\right|^2}{\epsilon_{\alpha}}~.
\end{equation}
Note that since the units of $\Gamma_{i\alpha}$ are
inverse length, $m_v$ indeed has the dimension of mass.
The prime on the summation implies omission of the zero modes, as
required by the constraint.  The vortex equation of motion
corresponding to this action contains the Magnus force
\begin{equation}
{\bf F}_M = m_v \frac{d^2{\bf r}_v}{dt^2} = 2 \hbar \pi n_c \; \Big( \hat{z} \times \frac{d {\bf r}_v}{dt} \Big).
\end{equation}
where the effective magnetic field strength is related to the
Berry-phase coupling by the expression $B_{\rm eff}=2\hbar {\rm
Im}[\Gamma_{xy}]$.

The vortex mass can be estimated by approximating the environment
modes as vortices modulated by plane-waves.  This is a good
approximation as long as we are interested in low energy
fluctuations in which vortex motion induces slowly-varying
non-uniformities. The phase and density environmental fluctuations
modes\index{environmental modes} in the presence of the vortex are
then given respectively by
\begin{equation}
\label{eq:Feynman1}
 {\chi}_{\bf k}^\theta = \frac{i}{\sqrt {2 n_c V}}
 \left(
    \begin{array}{c}
     \phi_{v} \\
      -\phi^{*}_{v}
    \end{array} \right) e^{i {\bf k} \cdot {\bf x}}~,
\end{equation}
and
\begin{equation}
\label{eq:Feynman2}
 {\chi}_{\bf k}^\rho=\frac{1}{\sqrt {2 n_c V}}
 \left(
    \begin{array}{c}
     \phi_{v} \\
      \phi^{*}_{v}
    \end{array} \right) e^{i {\bf k} \cdot {\bf x}}~,
\end{equation}
where $V$
is the area of the system.
Since these modes are spread over the full area of the
superfluid, the corresponding eigen-energies can be approximated by
their values in the absence of a vortex: $\epsilon_{\bf k}^{\theta} = \epsilon_{\bf k} = \hbar^2 {\bf k}^2/2m$
and $\epsilon_{\bf k}^{\rho} \approx \epsilon_{\bf k} + 2 g n_c$.  When
$\bf {k}=0$, these reduce to global phase and density
modes\index{density mode} respectively.

We now use these approximations to evaluate the coupling of a vortex to its
environment.  Separating the condensate wave function in the presence of a
vortex into amplitude and phase factors ($\phi_v = |\phi_{v}| \, e^{i\theta}$) we have that
\begin{equation}
\partial_x \phi_v ({\bf x}) = i |\phi_{v}| e^{i\theta} \; \partial_x \theta({\bf x})
+  e^{i\theta} \; \partial_x |\phi_v({\bf x})|,
\end{equation}
from which it follows that the coupling coefficients are given by
\begin{equation}
\Gamma_{{\bf k}x}^{\theta}= \frac{-i}{\sqrt{2n_cV}} \int d {\bf x} \; e^{-i\bf{k}\cdot{\bf x}} \;
|\phi_{v}|  \partial_x |\phi_{v}|,
\end{equation}
and
\begin{equation}
\Gamma_{{\bf k}x}^{\rho}= \frac{i}{\sqrt{2n_cV}} \int d {\bf x} \; e^{-i\bf{k}\cdot{\bf x}} \;
|\phi_{v}|^2 \; \partial_x \theta.
\end{equation}
The vortex-core size provides the key length scale for vortex-condensate coupling.
Since the core size corresponds to the coherence length scale $\xi$ at which
kinetic and interaction energy scales balance ($\hbar^2/2m\xi^2 = g n_c$), coupling is expected to
be large only when $k \xi$ is small, where $k=|{\bf k|}$.
$\Gamma_{{\bf k}x}^{\theta}$ is induced by changes in wave-function amplitude which are non-zero
only inside the vortex core.  Choosing polar coordinates centered on the vortex position and approximating
$|\phi_{v}|$ by $\sqrt{n_c} (r/\xi)$ inside the vortex core and by a constant $\sqrt{n_c}$ outside the core
with $\partial_{x}r = -e^{i\bf{k}\cdot{\bf x}}(d/dk_x e^{-i\bf{k}\cdot{\bf x}})/ir$ we obtain that
\begin{eqnarray}
\label{eq:phase}
\Gamma_{{\bf k}x}^{\theta} &\simeq& -\frac{1}{\sqrt{2n_cV}} \frac{n_{c}}{\xi^2}\frac{d}{dk_{x}}\int_{{\bf x} \in {\rm core}} d {\bf x} e^{-i\bf{k}\cdot{\bf x}} \nonumber \\
&\simeq&  \frac{k_{x}}{\sqrt{2n_cV}}\frac{n_c \pi  \xi^2}{4}~,
\end{eqnarray}
for small $k \xi$.  For the coupling to density fluctuations we note that the dominant contribution comes
from outside the vortex cores and write
$\partial_x \theta = - \sin(\theta)/r$ as $e^{i\bf{k}\cdot{\bf x}}(d/dk_y e^{-i\bf{k}\cdot{\bf x}})/ir^2$ to obtain
\begin{eqnarray}
\label{eq:density}
\label{eq:gammarho}
\Gamma_{{\bf k}x}^{\rho} &\simeq& \frac{2\pi n_c }{\sqrt{2n_cV}} \; \frac{d}{dk_y} \int_{k\xi}^{\infty} dx \, \frac{J_0(x)}{x} \nonumber \\
&\simeq& \frac{- 2\pi n_c}{\sqrt{2n_cV}}  \frac{k_y}{k^2}.
\end{eqnarray}
The final form for Eq.~(\ref{eq:gammarho}) is valid for $k\xi$
small.

Since the coupling between the vortex and its environment must be dominated by modes
with $k \xi$ small we can use these results to estimate the contributions of both phase ($\theta$) and
density ($\rho$) fluctuations to the vortex mass.  Substituting Eq.(~\ref{eq:phase}) and
Eq.(~\ref{eq:density}) into
Eq.~(\ref{eq:massadiabatic}) and using $2\pi/\xi$ as an ultraviolet cut-off
for the $m_v^{\theta}$ evaluation, we find that
\begin{equation}
m_v = m_v^{\theta}+m_v^{\rho} \approx m (n_c \pi \xi^2) \; \left[
\frac{\pi^2}{32} + \frac{\ln(R/\xi)}{2} \right].
\label{eq:vortexmass}
\end{equation}
The prefactor in this expression is the total mass of all boson particles inside the
vortex core; the two factors inside square brackets are estimates of the $m_v^{\theta}$ and $m_v^{\rho}$
contributions respectively.
Viewing the spatial cutoff $R$ as the smaller of the system size and the
distance between vortices, we see that the vortex-mass will normally be dominated
by the second term inside brackets, which represents coupling
to density fluctuations, especially so when the fraction of the
system area occupied by vortex cores is small.  We comment later that the
adiabatic approximation provides a good description of vortex quantum mechanics only
when the logarithmic enhancement of the vortex mass due to the its $1/r$
phase tail plays a substantial role.  In general
both contributions to the vortex mass should be retained.

\subsection{Beyond the adiabatic limit}
\index{dissipative action}
We have shown that we can reproduce the commonly-employed
vortex-particle approximation by making an adiabatic approximation to
our effective action. The next step is to scrutinize the adequacy
of the adiabatic approximation.
We expand the fluctuating vortex and environment fields,
\begin{eqnarray}
c_{\alpha}(\tau) &=& \frac{1}{\sqrt{\hbar\beta}}\sum_{n}c_{\alpha,n}e^{-i\omega_n\tau}~; \nonumber \\
c^{*}_{\alpha}(\tau) &=&\frac{1}{\sqrt{\hbar\beta}}\sum_{n}c^{*}_{\alpha,n}e^{i\omega_n\tau}~; \nonumber \\
r_{v}^{j}(\tau)&=&\frac{1}{\sqrt{\hbar\beta}}\sum_{n}r_{v,n}^{j}e^{-i\omega_{n}\tau}~,
\end{eqnarray}
in Matsubara frequencies $\omega_n=\frac{2\pi n}{\hbar\beta}$. The
vortex action\index{quantum dissipative action} is derived by
integrating over $c_{\alpha,n}$ and  $c^{*}_{\alpha,n}$ as in the
adiabatic case. The resulting action
\[
S_{\rm eff} [{\bf r}_v]
  =\sum_{n}
    \pi \omega_n \hbar n_c \epsilon^{jk}
   r_{v,n}^j
   r_{v,n}^k
\]
\begin{equation}
   \label{eq:seff2}
+\frac{1}{2}\omega_n^2 m_{v}^{jk} (i \omega_n)
   { r}_{v,n}^{j} { r}_{v,n}^{k},
\end{equation}
is of the same form as the adiabatic approximation action except that the
mass becomes frequency-dependent:
\begin{equation}
\label{eq:massnonadiabatic} m_{v}^{jk} (i \omega_n)
\equiv \hbar^2 \sum_{\alpha',\alpha}K_{\alpha',\alpha}
(i\omega_n)
   \Gamma_{r_v^{j},{\alpha'}}\Gamma_{r_{v}^{k},{\alpha}}^{*}
\end{equation}
where
\begin{equation}
\label{eq:mass kernel}
{K}_{\alpha,\alpha'} (i \omega_n) =
[\epsilon_{\alpha}\delta_{\alpha',\alpha}-2i\hbar\omega_{n}\Gamma_{\alpha',\alpha}]^{-1}.
\end{equation}
The mass kernel captures both the renormalization of the effective mass
at larger frequencies and the dissipative coupling of the vortex to the condensate
environment.
The quantity
\begin{equation}
\Gamma_{\alpha^{'},\alpha}=\frac{1}{2}\int dV [\chi^{1
*}_{\alpha^{'}}({\bf x}) \chi^{1}_\alpha({\bf x})-\chi^{2
*}_{\alpha^{'}}({\bf x}) \chi^{2}_\alpha({\bf x})]
\label{eq:gamma}
\end{equation}
is the
Berry-phase coupling between environment fluctuations which we have ignored in the
adiabatic theory.

Approximating, as before, density and phase
fluctuations by plane waves we replace state labels
$\alpha',\alpha$ by momentum labels ${{\bf k}'},{{\bf k}}$. We
divide the Berry-phase coupling between fluctuations
$\Gamma_{{{\bf k}'},{{\bf k}}}$ into different categories. Using
Eqs.~(\ref{eq:Feynman1}) and (\ref{eq:Feynman2}) we find that
\begin{equation}
\Gamma_{{{\bf k}'},{{\bf k}}}^{\rho \theta} = \frac{i}{2} \delta_{{{\bf k}'},{{\bf k}}}
\; \; \Gamma_{{{\bf k}'},{{\bf k}}}^{\theta \rho}= -\frac{i}{2} \delta_{{{\bf k}'},{{\bf k}}},
\end{equation}
and that there is no coupling among phase fluctuations and density fluctuations.
In this approximation then, the matrix $K_{\alpha',\alpha} (i \omega_n)$ is given
by $K_{\alpha',\alpha} = K_{{{\bf k}}} (i \omega_n) \delta_{{{\bf
k}'},{{\bf k}}}$ with the matrix $K_{{{\bf k}}}$ given by
\[
K_{{{\bf k}}} (i \omega_n) = \left(
  \begin{array}{cc}
    \epsilon^{\theta}_\bk & -\hbar\omega_{n} \\
    \hbar\omega_{n} & \epsilon^{\rho}_\bk \\
  \end{array}
\right)^{-1}
\]
\begin{equation}
=\frac{1}{(\epsilon^{\theta}_\bk \epsilon^{\rho}_\bk
+\hbar^2\omega_{n}^{2})}\left(
               \begin{array}{cc}
                 \epsilon^{\rho}_\bk & \hbar\omega_{n} \\
                 - \hbar\omega_{n} & \epsilon^{\theta}_\bk \\
               \end{array}
             \right)~.
\end{equation}
To study damping, we perform a Wick rotation\index{Wick rotation}
$i\omega_n\rightarrow \omega+i0 \equiv \omega^+$. The poles of the
matrix elements $K_{{{\bf k}}}[\omega^+]$ are the Bogoliubov modes
with excitation energy $E_{\bf k}^{b} =(\frac{\hbar^2
k^2}{2m}(\frac{\hbar^2 k^2}{2m}+2gn_c))^{1/2} $.

We now derive approximate analytic results for the non-adiabatic vortex
action.
The key step is to sum over environment fluctuation modes to obtain an expression
for the frequency-dependent mass kernel $m_{v}^{jj}$ where $j\in {x,y}$:
\begin{widetext}
\begin{eqnarray}
\label{eq:kernel}
m_{v}^{jj}(i\omega_{n}) &=& 2 \hbar^2\sum_{{{\bf k}}} \left(
  \begin{array}{cc}
   {\Gamma_{x{{\bf k}}}^{\theta}}^{*}  & {\Gamma_{x{{\bf k}}}^{\rho}}^{*} \\
  \end{array}
\right) K_{{{\bf k}}}(i\omega_n)\left(
\begin{array}{c}
\Gamma_{x{{\bf k}}}^{\theta}\\
\Gamma_{x{{\bf k}}}^{\rho}  \\
\end{array}
\right) \nonumber \\
&=& 2 \hbar^2 \sum_{{{\bf
k}}}\frac{1}{E_{b}^{2}(k)-(i\hbar\omega_{n})^{2}}
[2gn_{c}|\Gamma_{x{{\bf k}}}^{\theta}|^{2}
+\frac{\hbar^2k^2}{2m}|\Gamma_{x{{\bf k}}}^{\rho}|^2
+\hbar\omega_{n} ({\Gamma_{x{{\bf k}}}^{\rho *}}\Gamma_{x{{\bf
k}}}^{\theta}- {\Gamma_{x{{\bf k}}}^{\rho}}{\Gamma_{x{{\bf
k}}}^{\theta *}}) ],
\end{eqnarray}
\end{widetext}
At small ${\bf k}$ the Bogoliubov excitation energy\index{Bogoliubov excitation
energy} takes its phonon
dispersion\index{phonon dispersion} limit, $E_{b}=\hbar
k(gn_{c}/m)^{1/2}=\hbar kc$, where $c$ is the speed of sound in the
superfluid.  Note that the final term in square brackets vanishes.

We proceed as in the adiabatic limit, separating the mass into
phase $m_{\theta}$ and density $m_{\rho}$ fluctuation contributions:
\begin{widetext}
\[
m_{\theta}=\hbar^2\sum_{{{\bf k}}}
\left[\frac{1}{E_{b}(k)-\hbar \omega
-i\delta}+\frac{1}{E_{b}(k)+\hbar \omega+i\delta}\right]
\frac{2gn_{c}|\Gamma_{x{{\bf k}}}^{\theta}|^{2}}{E_{b}(k)}
\]
\begin{equation}
\label{eq:Mrho} \simeq \frac{\pi gn_{c}^2\xi^4}{32 c^{4}}
\int_{2\pi c/R}^{2\pi
c/\xi}d\omega'\left[\frac{\omega'^2}{\omega'+\omega}+
P[\frac{\omega'^2}{\omega'-\omega}]+i\pi\omega^2\delta(\omega'-\omega)\right].
\end{equation}
\end{widetext}
Similarly, the condensate density fluctuation contribution $m_{\rho}$ is
\begin{widetext}
\begin{eqnarray}
m_{\rho}&=&2 \hbar^{2}\sum_{\bk}\frac{1}{E_{b}(k)}
\left[\frac{1}{E_{b}(k)-\hbar \omega
-i\delta}+\frac{1}{E_{b}(k)+\hbar \omega+i\delta}\right]
\frac{\hbar^2 k^2}{4m}|\Gamma_{x{{\bf k}}}^{\rho}|^{2}  \nonumber \\
&\simeq& \frac{m \pi\xi^2 n_{c}}{4} \int_{\omega}^{2\pi
c/\xi}d\omega'\left[\frac{1}{\omega'+\omega}+
P[\frac{1}{\omega'-\omega}]+i\pi\delta(\omega'-\omega)\right] \;
\simeq \; \frac{m \pi\xi^2n_{c}}{4} \; [2 \ln(2\pi c/\omega \xi) +
i \pi ].
\end{eqnarray}
\end{widetext}
Assuming that the density-coupling is dominant we see that the
dissipative contribution to the mass is frequency-independent and
that the real part of the mass has a corresponding logarithmic
dependence on frequency.  This result is similar to that of Arovas
and Freire \cite{arovas1997} obtained by different methods.
The adiabatic limit result for the mass quoted previously
corresponds to setting the frequency argument in this expression
to its minimum value $\sim c/R$.
The Quality-factor for vortex cyclotron motion in the adiabatic limit is therefore
\begin{equation}
(\omega_c \tau)_{vortex} \;  \sim \;\; \ln(R/\xi).
\end{equation}
Sharp cyclotron modes are expected only for very dilute vortices or for
isolated vortices in a large condensed cloud.

We can now comment on the validity of using the adiabatic approximation
to describe vortex quantum fluctuations.
As a criterion
for the validity of the adiabatic theory in the continuum limit,
we require that the adiabatic cyclotron frequency $\omega_c$ be
much smaller than $c/\xi$, that is,
\begin{equation}
\omega_{c} = 8\frac{\hbar}{m\xi^2 \ln{\frac{R }{\xi}}} \ll \frac{c}{\xi} \simeq \frac{\hbar}{m\xi^{2}}.
\end{equation}
This inequality is satisfied only if $R$ or the mean-distance between vortices is at least
a thousand times $\xi$, {\em i.e.} only if an extremely small fraction of the system area
is occupied by vortex cores.  This condition seems to be quite difficult to satisfy for a
cold-atom bosonic system.  In the next section we explore the possibility of
altering this comparison using the condensate tuning possibilities available
in optical lattices.

\section{Vortex Cyclotron Modes in Optical Lattices}
\label{sec:OL}
\subsection{Boson Hubbard Model}
Cold atoms in optical lattices offer the opportunity to alter
vortex physics by tuning system parameters like the interaction
strength between atoms, and the optical potential depth and
period. When an optical lattice\index{optical lattices} potential
is sufficiently deep, low-energy physics is described by a one
band Hubbard model\cite{Greiner,RMP} This standard Boson Hubbard
model\index{Bose-Hubbard model} without an external trapping
potential is described by the Hamiltonian
\begin{equation}
\label{eq:Hubbard model} H_{BH}= -J\sum_{\langle
i,j\rangle}b_{i}^{\dagger}b_{j}
+\frac{U}{2}\sum_{j}b_{j}^{\dagger}b_{j}^{\dagger}b_{j}b_{j}-\mu
\sum_{j}b_{j}^{\dagger}b_{j}.
\end{equation}
Here $\langle i,j\rangle$ represents summation over the nearest
neighbors, $J$ is the tunneling amplitude and $U$ is the
interaction energy between atoms on the same site, and $\mu$ is the chemical potential\index{chemical
potential}.
The energy functional ${\mathcal E}$ in the coherent state path
integral action for the Boson Hubbard model is correspondingly
given by
\begin{equation}
E_{SF}=-J\sum_{<i,j>}\phi_{i}^{*}\phi_{j}
+U\sum_{j}\phi_{j}^{*}\phi_{j}^{*}\phi_{j}\phi_{j}-\mu
\sum_{j}\phi_{j}^{*}\phi_{j}.
\end{equation}
Mean-field configurations
of the condensate are extrema of this energy functional and therefore
satisfy the corresponding Gross-Pitaevskii (GP) equation.
The Boson Hubbard model on a 2D square lattice with lattice constant $D$ can be viewed
as a discretization of the 2D continuum model we have been discussing
heretofore with the correspondences
\begin{equation}
\label{eq:mapping1}
J\leftrightarrow\frac{\hbar^2}{2mD^2}
\end{equation}
 and
\begin{equation}
\label{eq:mapping2}
U \leftrightarrow
\frac{g}{D^{2}}.
\end{equation}
Based on continuum model considerations explained earlier, we expect the
adiabatic approximation to be satisfied when the ratio of the number of
lattice sites per vortex to the number of lattice sites occupied by a vortex core
is large.  In searches for vortex-Landau quantization then, the small vortex
cores produced by strong interactions are favored.  On the other hand we know that
the Bose condensate is destroyed by quantum fluctuations for\cite{bhm_sit}
$U > U_{c} \simeq 4 J z \bar{n}$ where $z$ is the lattice coordination number
and $\bar{n}$ is the average number of bosons per lattice site.
($z=4$ in our case.)  Using the continuum model correspondences
we estimate that
\begin{equation}
\frac{\xi^2}{D^2} \sim \frac{J}{U} \sim \frac{U_c}{U} \frac{1}{4z\bar{n}}.
\end{equation}
There should therefore be a broad range of interaction strengths over which
superfluidity survives and vortex cores are concentrated in a single
square lattice plaquette, even for the small values of $\bar{n}$ typical in
optical lattice systems.  The numerical calculations reported on below
explore the possibility, suggested by the approximate continuum model calculations,
that sharp adiabatic-limit vortex cyclotron motion modes might occur
for $U$ strong, but still smaller than $U_{c}$.

\subsection{Vortex Displacement Fluctuations}

In this section we describe numerical calculations based on the
theoretical formulation outlined in Section \ref{sec:theory},
which demonstrate the possibility of achieving some degree of
experimental controls over the sharpness of the vortex cyclotron
resonance signal.  To find solutions of the square lattice Hubbard
model Gross-Pitaevskii equations which contain vortices, we impose
quasi-periodic boundary conditions on a finite lattice with size
$L \times L$. We first impose a phase winding of $2\pi$ on this
area by using the boundary conditions
\begin{eqnarray}
\psi(x,\frac{L}{2}) &=& e^{i\pi(1-\frac{x}{L})}\psi(x,-\frac{L}{2}), \nonumber \\
\psi(\frac{L}{2},y)&=& e^{-i\pi(1-\frac{x}{L})}\psi(-\frac{L}{2},y),
\end{eqnarray}
where the origin is chosen at the center of the computational grid
which is also the center of the vortex core.
We then minimize the energy by integrating the time-dependent Gross-Pitaevskii
equations over imaginary time,

\begin{equation}
\psi_{i,j}^{n+1}-\psi_{i,j}^{n}= - \epsilon \, H_{GP} \, \psi_{i,j}^{n},
\end{equation}
 carefully
restoring the normalization $\int d^{2}{\bf
x}|\psi_{i,j}^{n+1}|^{2}=N_{a}$ after each time step and starting
from an ansatz in which the phase winds at a constant rate around
the boundary.  Here $i,j$ are labels for the two-dimensional
lattice points, the label $n$ counts steps in an imaginary time
evolution with time-step $\epsilon$ and $N_{a} =  \bar{n} L^2/D^2$
is the number of atoms in the $L \times L$ computational cell.

The energy functional for fluctuations around the vortex ground
state are described by the lattice version of the Hermitian
Hamiltonian in Eq.~(\ref{eq:hamflucs}). Diagonalizing this
Hamiltonian, we always find a zero-energy global phase mode and
two other low but finite energy fluctuation modes which we
identify as vortex translational modes.  The vortex modes do not
have precisely zero-energy in our simulations because of lattice
pinning and because of the boundary conditions we apply to the
simulation cell.  The number of eigenvalues of $\cal{H}$ is twice
the number of atoms in the simulation cell. In order to connect
approximately with an infinite system with a finite density of
vortices, we broaden this discrete spectra into continua in the
spectral calculations described below
by replacing delta-functions at eigenenergies by
Lorentzian functions with a width $\delta \sim \hbar^2/2mL^2$.

We report results for four different interaction strengths,
$U=0.05J$, $U=0.40J$, $U=2.25J$ and $U=3.80J$ which we refer to
below as very weak, weak, strong, and very strong.  According to
our estimates only the very weak and weak cases should yield
vortex cores which extend over many square lattice plaquettes and
this expectation is confirmed in Fig.~\ref{Pedagogical}. All our
calculations were performed for a square lattice system with one
vortex per $N_{site}= 46 \times 46 = 2116$ lattice sites. In order
to compare with potential future optical lattice experiments we
choose the number of atoms per site $\bar{n}=1$ in all the
calculations described below.  Our approach, which is designed
to describe systems with weak and moderately strong interactions,
is complementary to that
taken by Lindner {\em et al.}\cite{auerbach} who use many-body
exact diagonalization methods to estimate vortex properties and concentrate
on the case of a half-filled lattice with infinitely strong on-site
repulsive interactions.

\begin{figure}
\centering
\includegraphics[width=3.5in]{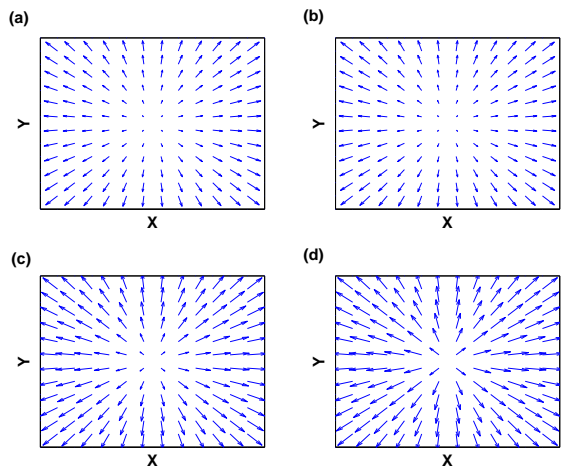}\\
  \caption{Condensate amplitude and phase for the vortex-state configuration indicated by the length and
orientation of an arrow on a set of $12 \times 12$ sites near the center of the $46 \times 46$ site simulation cell:
a) $U/J =0.05$ b)  $U/J=0.40$, c) $U/J=2.25$ and  $U/J=3.8$.
The continuum model estimate for the vortex core size in square lattice constant units is
 $\xi = D (J/U)^{1/2} =4.4721D, 1.5811D, 0.6667D, 0.5130D$
for cases (a), (b), (c) and (d) respectively.}
  \label{Pedagogical}
\end{figure}

The two degenerate vortex translation modes are identified as
$c_{x}(\tau)\delta \phi_{0x}$ and $c_{y}(\tau)\delta \phi_{0y}$,
corresponding to vortex displacements along the x-axis and y-axis
of the lattice respectively.  Here $\delta \phi_{0i}$ is a
normalized eigenfunction of the system energy functional and is
obtained numerically. To make contact with the vortex coordinate
${\bf r}_v= (x_{v}(\tau),y_{v}(\tau))$ in our continuum field
theory, we can associate the excitation amplitudes $c_{x},c_{y}$
with the vortex coordinates $x_{v},y_{v}$ using the relations
\begin{eqnarray}
c_{x}\delta \phi_{0x} &=& - x_{v}\partial_{x} \phi_{v} \nonumber \\
c_{y}\delta \phi_{0y} &=& - y_{v}\partial_{y}\phi_{v}.
\end{eqnarray}
This identification becomes exact in the continuum limit where
$\phi_v({\bf x} -{\bf r}_{v}(\tau))=\phi_{v}(\bf x)-{\bf
r_{v}}\cdot \nabla \phi_{v}$ for small vortex displacements. In
this limit $c_{x}$ and $c_{y}$ represent precisely the
same degrees of freedom as $x_{v}$ and $y_{v}$ and differ only by
a multiplicative constant:
\begin{equation}
\label{normalization} \frac{x_{v}}{c_{x}} = \frac{y_{v}}{c_{y}} =
- \Big[\frac{1}{\int d{\bf
x}|\partial_{x}\phi_{v}|^{2}}\Big]^{1/2}.
\end{equation}
In the continuum limit
\begin{equation}
\label{eq:renorm} \frac{x_{v}}{c_{x}}=\int d{\bf
x}(f^{2}(\partial_{x}\theta)^{2}+(\partial_{x}f)^{2})^{-1/2}
\approx [\pi n_{c}(1+\ln{R/\xi})]^{-1/2}~.
\end{equation}

When we turn to the Berry-phase coupling coefficients for the
lattice model fluctuations below, they are first evaluated using
Eq.~(\ref{eq:gamma}) with the numerical eigenstates of the energy
functional. Applying a central difference approximation,
$\partial_x\phi_{v}(i,j) \simeq
[\phi_{v}(i+1,j)-\phi_{v}(i-1,j)]/2D$, to the vortex solutions of
our lattice model ($i,j$ are the lattice point indices),
Eq.~(\ref{normalization}) then allows us to appropriately
renormalize those coupling coefficients that involve
vortex-displacement coordinates. This identification of particular
fluctuation modes as vortex displacements is necessary in order to
evaluate the adiabatic vortex mass using
Eq.~(\ref{eq:massadiabatic}), or its frequency-dependent
non-adiabatic generalization using
Eq.~(\ref{eq:massnonadiabatic}), but drops out of any direct
calculation of a physical observable. In particular the
renormalization does not influence the frequency or broadening of
the mode which we identify as a vortex cyclotron resonance. Its
only role is related to the physical identification of a
particular low-frequency collective fluctuation coordinate as the
position of a well-defined texture of the condensate field.  The
identification becomes less well founded, of course, as the vortex
cores become smaller and their locations more strongly pinned by
the lattice.  We will return to this point frequently below.

\begin{figure}
\centering
\includegraphics[width=3.5in]{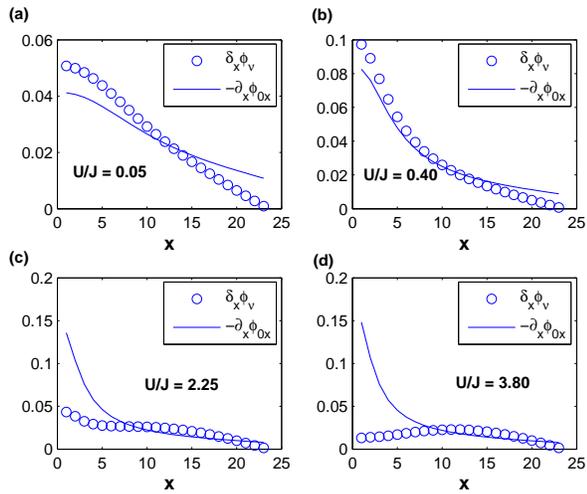}\\
  \caption{Comparison of the lattice-model energy functional eigenstate identified as a vortex displacement in the $\hat{x}$ direction
($\delta \phi_{0x} ({\bf x})$ - circles) with the renormalized
finite-difference derivative of the lattice-model vortex
wavefunctions (solid line) illustrated in Fig.~\ref{Pedagogical}
for the same four values of $U/J$.  Distance in this figure is in
units of the lattice constant $D$ of the optical lattice.}
  \label{fig:FIG4}
\end{figure}

In Fig.~\ref{fig:FIG4}, we compare the normalized vortex
displacement mode $\delta \phi_{0x}({\bf x})$ obtained by
diagonalizing the energy functional in Eq.~\ref{eq:hamflucs} with
the renormalized finite-difference approximation to the
derivative, $-\partial_{x}\phi_{v}$, of the vortex condensate
configurations illustrated in Fig.~\ref{Pedagogical}. For the weak
interaction cases (Figs.~(\ref{fig:FIG4}a) and (\ref{fig:FIG4}b))
the close comparison between the two functions clearly justifies
the identification. In Fig.~(\ref{fig:FIG4}a), the vortex core
size is not very much smaller than our simulation cell size and we
attribute differences between the two-functions to boundary
effects. In case (\ref{fig:FIG4}b), the vortex core is smaller and
boundary effects are less important. In the strong interaction
cases, (\ref{fig:FIG4}c) and (\ref{fig:FIG4}d), we see the
influence of lattice pinning which suppresses the numerical
fluctuation mode close to the vortex core, more so for stronger
interactions.  The ratio between $\delta \phi_{0x}({\bf x})$ and
$-\partial_{x}\phi_{v}$ implied by the analytical estimate,
Eq.~(\ref{eq:renorm}), varies from $\sim 0.31 D$ in the weak
interaction limit to $\sim 0.24 D$ for the strong interaction
limit and is in good qualitative agreement the scaling factors
used to normalize $-\partial_{x}\phi_{0x}$ in Fig.~\ref{fig:FIG4}.

\subsection{Vortex Effective Mass}

\begin{figure}
\centering
\includegraphics[width=3.8in]{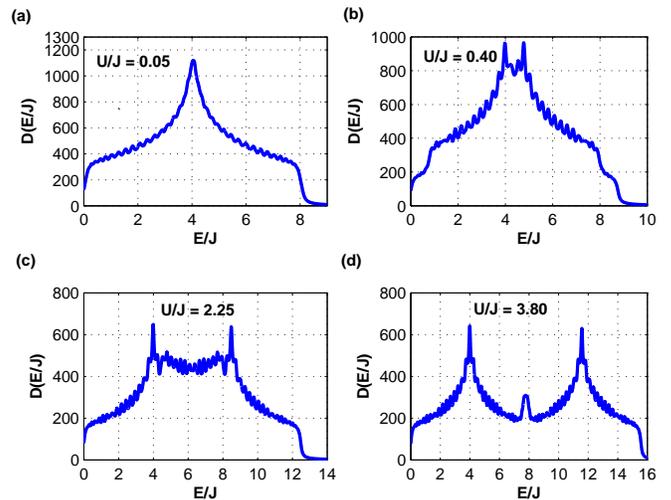}\\
  \caption{Density of states $D(E_{R})$ of environment fluctuation modes
  as a function energy $E/J$ for $U/J=0.05$ (a), $U/J=0.40$ (b), $U/J=2.25$ (c), and $U/J=3.80$ (d).
  }\label{fig:FIG1}
\end{figure}

As explained earlier, the vortex effective mass is due to coupling between the vortex displacements
and their environment which, in the Gaussian approximation we employ here, consists of
condensate phase and density fluctuations.  In Fig.~\ref{fig:FIG1}, we plot the density-of-states for condensate density and phase fluctuation modes,
distorted by the presence of the vortex.  In the absence of a vortex we would expect that
the phase modes should have energy $\epsilon^{\theta}_{{\bf k}} = 2J(2-\cos(k_xa)-\cos(k_ya))$, varying over
an $8J$ range, and that the density fluctuation modes have energy $\epsilon^{\rho}_{{\bf k}} = \epsilon^{\theta}_{{\bf k}}+ 2 U$.
We see in Fig.~\ref{fig:FIG1} that the presence of the vortex does not have a large
influence on the overall spectrum of fluctuations.  The prominent features at
$E=4J$ and $E=4J+2U$ in Fig.~\ref{fig:FIG1} originate from the van-Hove singularities of the
nearest-neighbor hopping model on the square lattice.
As the interactions strengthen, the phase and density
fluctuation spectra separate.
Note that the densities of states differ by a factor of two
between case (a) in which the phase and density spectra overlap
and case (d) in which they are nearly completely separated.

\begin{figure}
\centering
\includegraphics[width=3.8in]{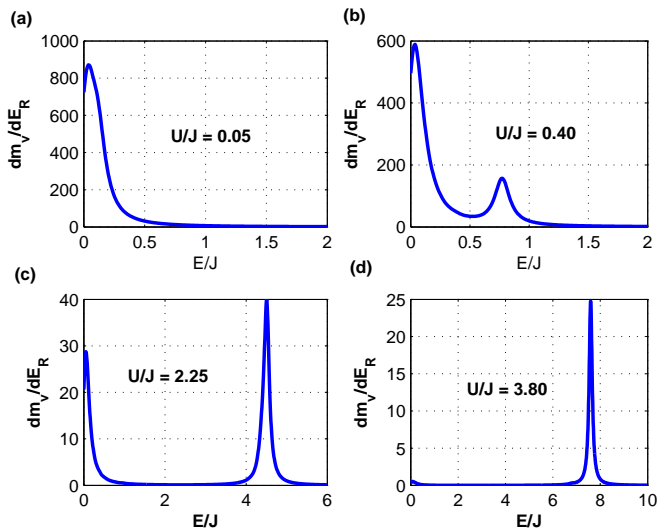}
\caption{Spectral representation of adiabatic-vortex-mass contributions $dm_v/dE$
{\em vs.} mode energy $E/J$ for
$U/J=0.05$ (a), $U/J=0.40$ (b), $U/J=2.25$ (c), and $U/J=3.80$ (d).
Where a separation exists, the lower-energy contribution can be associated
with coupling to phase fluctuations and the higher energy contribution with
coupling to density fluctuations.
The density-fluctuation contribution has a larger relative importance when
interactions are strong.}\label{fig:FIG2}
\end{figure}

In Fig.~\ref{fig:FIG2} we plot a spectral decomposition of adiabatic-vortex-mass
contributions $dm_v/dE$ defined by
\begin{equation}
m_{v}= \hbar^{2}\sum_{\alpha}\frac{|\Gamma_{\alpha
x}|^{2}}{\epsilon_{\alpha}} \equiv \int dE \;  \frac{dm_v}{dE}~.
\label{eq:lattice_mass}
\end{equation}
In the very weak interaction case (Fig.~\ref{fig:FIG2}(a)) density and phase fluctuation
contributions are not spectrally resolved.  We see here that the mass is due
mainly to low-energy long-wavelength fluctuation modes.  For Figs.~\ref{fig:FIG2}(b-d)
we see separate contributions near $E=0$ due to phase fluctuations and
near $E=2U$ due to density-fluctuations.  As the interaction strengthens
and the vortex-core radius gets smaller, the density-fluctuation contribution
becomes relatively more important, in agreement with our continuum model
analytic results.  These results emphasize the fact that for moderate vortex
densities, both phase and density fluctuation contributions need to be included in
calculating the vortex mass.

\begin{figure}
\centering
\includegraphics[width=3.8in]{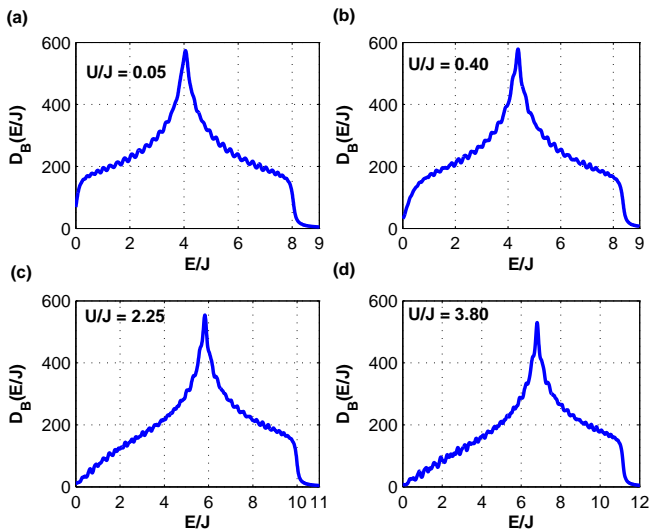}\\
  \caption{Spectral density of states of Bogoliubov excitations $D_{B}(E)$ as a function of
  $E/J$ for interaction strengths $U/J=0.05$ (a), $U/J=0.40$ (b), $U/J=2.25$ (c), and $U/J=3.80$ (d).
  The Bogoliubov modes have linear dispersion at long-wavelengths and a corresponding
  linear density of states at low energies.  At higher energies their
  density of states shows the characteristic van-Hove singularities of square lattice systems
  with near-neighbor hopping.  Linear dispersion holds over a wider energy range when the interactions are
  stronger.
  }\label{fig:FIG3}
\end{figure}

When we go beyond the adiabatic approximation, the kinetic coupling between
environment phase and density fluctuations is retained.  This coupling leads
(in the Gaussian fluctuation approximation) to Bogoliubov elementary-excitations
of the environment.  In the absence of a vortex, the environment fluctuations can
be labeled by wavevector ${\bf k}$ and the elementary excitation energy is given by
$E_{B}({\bf k}) = ( \epsilon^{\theta}_{\bf k} \, \epsilon^{\rho}_{\bf k} )^{1/2}$,
with characteristic linear dispersion at long wavelength.
For each ${\bf k}$ the elementary-excitation energy is therefore intermediate between the
energies of the corresponding phase and density modes.  There is now a single branch of modes,
instead of two-different ones.  As shown in the approximate continuum model calculations,
the phase and density contributions appear separately; their
correlation functions have different residues at elementary-excitation poles.
The mass becomes $\omega$-dependent and develops an imaginary part which captures the dissipative effect of environmental coupling.
The sign of the contribution of a particular elementary excitation to the mass changes when $\omega$ becomes larger than its energy.

In Fig.~\ref{fig:FIG3} we plot the density of states $D_{B}(E)$ of Bogoliubov
excitations \index{density of states of Bogoliubov
excitations}in the presence of a vortex, for the usual set of interaction strengths.
It is clear that distortion by the vortex has little influence on such global
indicators of environment properties.  For weak interactions the density-of-states
is roughly constant except at very low energies, corresponding to quadratic dispersion
at small ${\bf k}$, whereas for stronger interactions the density-of-states has
a linear dependence of energy, corresponding to linearly dispersing phonon-like
elementary excitations.

Having characterized the environment modes, we now discuss the adiabatic vortex mass $m_{v}$ and Berry phase $\Gamma_{xy}$ for
optical lattice condensates.  As explained earlier both of these quantities
are sensitive to the conversion factor
between the vortex coordinate $x_{\nu}$ and the excitation amplitude of the
corresponding normalized energy fluctuation mode.
The procedure we use to define this conversion factor, explained above, becomes somewhat arbitrary when interactions are very
strong and vortices are strongly pinned by the lattice.  The vortex cyclotron frequency, $\omega_{c}=2\hbar {|\Gamma_{xy}|}/m_{v}$,
is however independent of this conversion factor.  This expression for $\omega_{c}$ follows from the effective action in Eq.~(\ref{eq:seff2}) after Wick
rotation to real frequencies using an adiabatic-approximation frequency-independent vortex mass.

\begin{figure}
\centering
\includegraphics[width=3.2in]{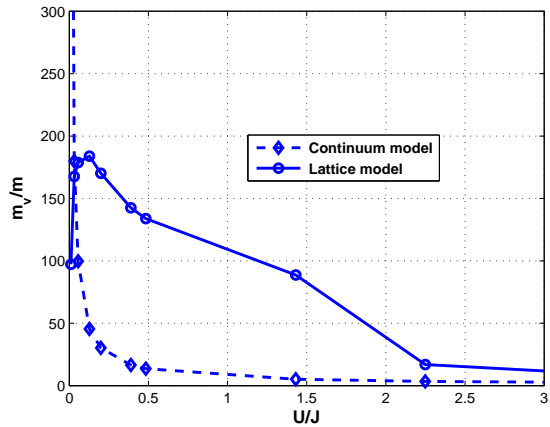}\\
  \caption{Vortex mass $m_{v}$ as a function of interaction strength $U/J$. The dashed-diamond line is a plot of the
 the analytical continuum model expression, Eq.~(\ref{eq:vortexmass}), using the continuum-limit expression for the
 vortex-core size $\xi$.  The solid-circle line shows vortex masses evaluated numerically from Eq.~(\ref{eq:lattice_mass}).
The numerical calculation is dependent on the conversion factor between the
    excitation amplitudes $c_{x}(\tau)$ and $x_{\nu}(\tau)$ as discussed in the text.}     \label{fig:FIG5}
 \end{figure}

In Fig.~\ref{fig:FIG5},
we illustrate the dependence of the adiabatic vortex mass on the interaction strength $U/J$.
When interactions are strong and vortex cores are small ($U/J \gtrsim 1$), we see that the vortex mass is not much larger than the bare particle mass.
In this strong interaction regime our numerical adiabatic vortex masses agree reasonably well with the continuum model expression translated
into lattice model parameters:
\begin{equation}
\frac{m_v}{m} \simeq \frac{\pi \bar{n} J}{U} \, \left[ \frac{\pi^2}{32} + \frac{\ln{N}}{4} + \frac{\ln(U/J)}{4} \right]
\end{equation}
where $N=2116$ is the number of lattice sites in our simulation cell.
This expression is also plotted in Fig.~\ref{fig:FIG5}.    There is a reasonable degree of
agreement between approximate analytic and numerical masses
in the strong interaction limit, even when
the vortex core size becomes comparable to $D$, possibly because mass contributions then come
mainly from interactions with density-fluctuations outside the
vortex core.  At weaker interactions, where the vortex cores become large, the deviations from the analytic expression are
larger.  In particular for very weak interactions the analytic expression has $m_{v}$ increasing as $U^{-1}$, a
trend that is not reproduced by our numerical calculations.  We attribute this discrepancy to
a violation in the numerical calculations of the assumption made in the analytic calculations that the vortex core size is
small compared to $R$.  At intermediate interaction strengths there is reasonable quantitative
agreement between the analytic approximation, which is relatively crude when applied to a lattice
model, and the numerical adiabatic mass results.

\begin{figure}
\centering
\includegraphics[width=3.2in]{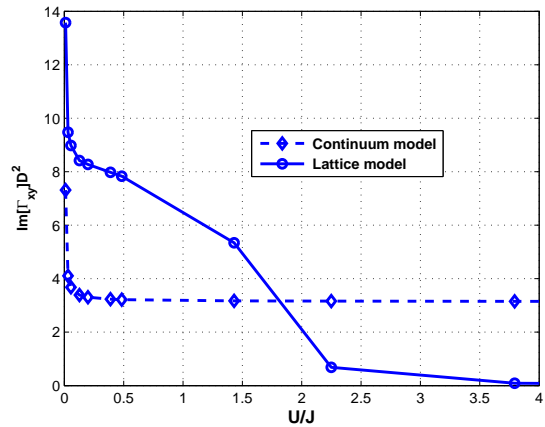}\\
  \caption{Berry-phase coupling ${\rm Im}[\Gamma_{xy}]=|\Gamma_{xy}|$ between $x_{v}$ and
$y_{v}$ as a function of interaction strength $U/J$. The dashed-diamond line is the data calculated from Eq.~(\ref{eq:Berryxy}) for the continuum model.
The upturn at small $U/J$ reflects the fact that the condensate density evaluated at the simulation cell boundaries is
increased compared to the average condensate density when the vortex cores occupy a significant fraction of the simulation
cell area.
  The solid-circle line is the data for the lattice model directly calculated from the wave function of the vortex displaced modes $\delta \phi_{0x}, \delta \phi_{0y}$.}
  \label{fig:FIG6}
\end{figure}

Next we discuss numerical results for $\Gamma_{xy}$,
the Berry-phase coupling between $x_v$ and $y_v$.  This coupling is the source of the
effective magnetic field $B_{eff}$ responsible for vortex cyclotron resonance.
In the continuum model $\Gamma_{xy} = \pi n_{c} \to \pi \bar{n}/D^2 \to \pi/D^2$.
We see in Fig.~\ref{fig:FIG6} that the numerical lattice model $\Gamma_{xy}$ is in
agreement only at intermediate values of $U/J$.  We attribute deviations at small $U/J$ to
the finite size of our simulation cell, as in the adiabatic effective mass discussion.
However, $\Gamma_{xy}$ also deviates strongly from the analytic result at large values of $U/J$.
We attribute this to the fact that vortex cores in this regime are localized mainly in the central plaquette
of the lattice. As a result, the condensate density variation is almost negligible across the entire lattice.
Since the continuum model $\Gamma_{xy}$ is proportional to the difference between
the condensate density far away from and at the center of the vortex core, the
small values obtained for $\Gamma_{xy}$ in this limit are not surprising.
The strong discrepancy compared to the continuum model in this regime is also due to lattice pinning effects
which work against the identification of the low-energy fluctuation modes with vortex translations.

\begin{figure}
\centering
\includegraphics[width=3.2in]{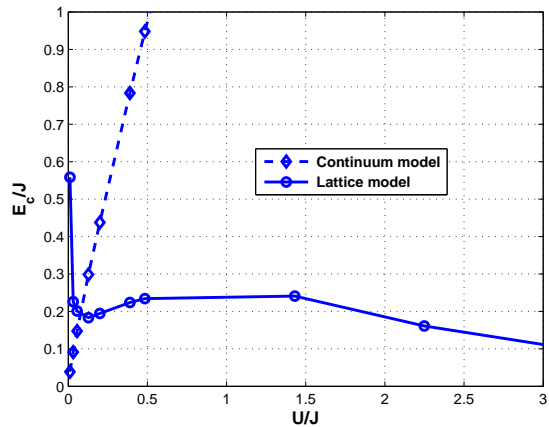}\\
  \caption{Vortex cyclotron frequency $E_{c}/J$ as a function of interaction $U/J$.  The dashed line shows the
continuum model estimates while the solid line illustrates numerical lattice model results.}
  \label{fig:FIG7}
\end{figure}

Before turning to the full non-adiabatic calculations, we first discuss
the adiabatic approximation for the vortex cyclotron frequency in the lattice model
which is proportional to $\Gamma_{xy}/m_v$.
Using Eqs.~(\ref{eq:vortexmass}), (\ref{eq:mapping1}), and (\ref{eq:mapping2}),
the continuum model adiabatic cyclotron mode energy translates into the following
lattice model expression:
\begin{equation}
E_{c} \simeq  \frac{ 8 \bar{n}U }{\frac{\pi^2}{16}+ \frac{\ln{N}}{2} + \frac{\ln(U/J)}{2}}
\end{equation}

Based this expression, we should expect that
the cyclotron mode frequency $E_{c}/J$ would scale with the interaction strength ${\bar n} U/J\rightarrow U/J$,
as indicated by the dashed line in Fig.~\ref{fig:FIG7}.  The discrepancies between continuum model and
lattice model results illustrated in Fig.~\ref{fig:FIG7} are mainly due to discrepancy in $\Gamma_{xy}$ values
at large $U$ and due to large vortex core sizes and related discrepancies in the vortex mass at small $U$.  At intermediate values of
$U$ the numerical lattice model adiabatic approximation and analytic continuum model
adiabatic approximation give similar values for the
vortex cyclotron frequency.

\subsection{Vortex Cyclotron Motion}

In Fig.~\ref{fig:FIG8}, we show the behavior of the real and
imaginary parts of the frequency-dependence in
$m_{v}^{xx}=m_{v}(E=\hbar\omega)$ for various values of $U/J$. The $\omega
\to 0$ limit of $m_{v}(E=\hbar\omega)$ is the adiabatic vortex
mass discussed above.  The contribution of particular environment
Bogoliubov modes to the vortex mass increases in magnitude as
$\hbar \omega$ approaches the energy of that mode from below and
then changes sign at higher frequencies.  The mass contribution
vanishes for $\omega$ much larger than environment mode
frequencies.

The real and imaginary parts of the vortex mass are related by a
standard Kronig-Kramers relationship.  [See
Eq.~(\ref{eq:kernel}).] We see in Fig.~\ref{fig:FIG8} that for the
weak interaction cases the imaginary part of the vortex-mass
approaches a constant at small frequencies, as in the continuum
case and that the real part of the vortex mass has a corresponding
logarithmic peak at low frequencies.  At strong interactions the
long-wavelength low-energy contribution to ${\rm
Im}[m_{v}(E=\hbar\omega)]$ is not dominant but a clear peak
appears at the energy of the van-Hove singularity in the
Bogoliubov spectrum. This peak becomes quite strong for strong
interactions and the real part of the vortex mass correspondingly
becomes quite large before changing sign near this frequency.
Because the imaginary part of the vortex mass is relatively small
and the real part is relatively frequency independent over a broad
frequency regime, we conclude that the adiabatic approximation
might be reasonable in describing the lattice-pinned vortex degree
of freedom in strongly interacting optical lattices.  Of course,
as we have already explained, continuum model estimates completely
fail to describe the frequency of the mode in this regime.

\begin{figure}
\centering
\includegraphics[width=4in]{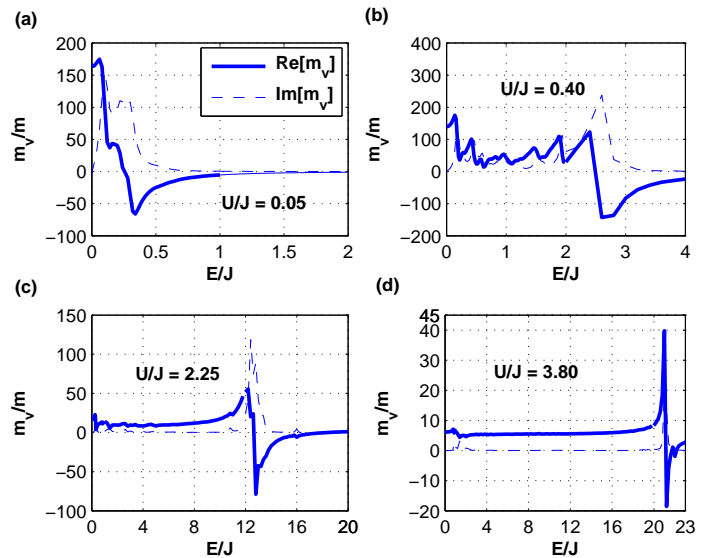}\\
  \caption{Real part and imaginary part of the frequency dependent vortex mass $m_{v}(E)$ as a function of the energy in units of $J$ for
the four different interaction strengths.  The irregular frequency
dependence at low frequencies in some plots is simply a reflection
of incomplete broadening of our discrete Bogoliubov excitation
spectrum and is without physical significance.}
  \label{fig:FIG8}
\end{figure}

In Fig.~\ref{fig:FIG9}, we test for the existence of a sharp
cyclotron mode by examining the frequency dependence of the
imaginary part of the vortex position-position spectral function
$A_{0i}[\omega] = -\frac{1}{\pi}{\rm Im} G^{R}_{0i}$ where
$G^{R}_{0i} = \langle x_{\nu}(\omega)x_{\nu}(-\omega)\rangle =
\langle y_{\nu}(\omega)y_{\nu}(-\omega)\rangle$. The corresponding
adiabatic approximation in the lattice model is obtained by
setting the off-diagonal elements between environmental modes
$\Gamma_{\alpha,\alpha'}$ to zero in our calculations. As
expected, the adiabatic vortex picture is reasonably accurate for
the strongly interacting cases.  For the weak interaction case
illustrated in Fig.~(9a), the vortex cyclotron mode frequency is
renormalized and is strongly broadened because ${\rm
Im}[m_{v}(E)]$ is large at the resonance position. The adiabatic
cyclotron modes in cases (b) and (c) have a complicated frequency
dependence because of the combined influence of long-wavelength
and van-Hove singularity environment modes. In Figs.~(9d), the
high-frequency modes near the van-Hove singularity are strongly
dominant and corrections to the adiabatic approximation are weak.

\begin{figure}
\centering
\includegraphics[width=3.8in]{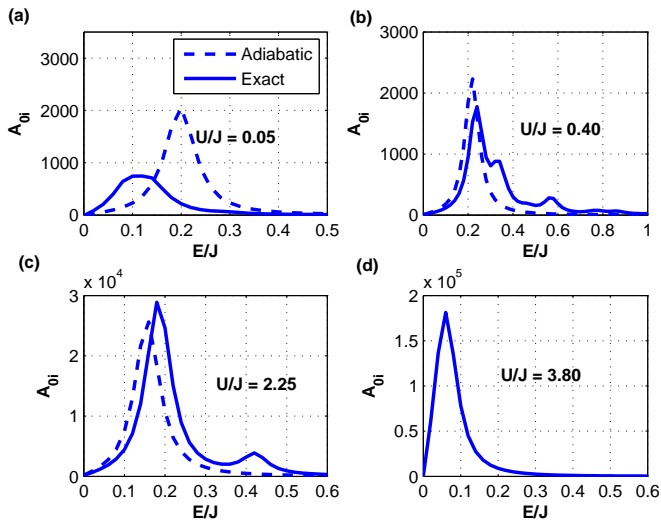}\\
  \caption{Imaginary part of vortex position-position correlation function $A_{0i}[\omega]$ with (dashed lines) and without (solid lines)
  the adiabatic approximation.   The adiabatic approximation becomes more accurate
  when interactions are very strong (case (d)) because the lattice-weakened Berry-phase coupling $\Gamma_{xy}$
suppresses the resonance frequency.  The adiabatic approximation result for $U/J = 3.80$ coincides closely
with its non-adiabatic counterpart. }\label{fig:FIG9}
\end{figure}

\section{Conclusions and discussion} \label{sec:conclusion}\

In this paper we have developed an approach to vortex dynamics in
neutral Bose superfluids that is based on careful separation of
zero modes corresponding to vortex translation from the other
fluctuations around the mean-field Bose condensate. Integrating out other
fluctuations leads to a vortex action with a frequency-dependent vortex mass.
In the adiabatic limit, in which frequency dependence is ignored, we recover
the result of Duan \cite{Duan1993} that the vortex mass diverges
logarithmically with system size in a continuum model of the
superfluid. The frequency dependence of the effective mass found
in our approach is similar to the result obtained by Arovas and Freire
\cite{arovas1997} with a completely different method. The
frequency-dependent effective mass has a nonzero imaginary part
which captures the dissipation associated with vortex motion.

In our approach Berry-phase coupling between orthogonal vortex displacements,
which causes the vortex to behave like a charged particle in an effective
magnetic field, can be calculated straightforwardly. In the adiabatic limit
the vortex undergoes undamped cyclotron motion which
quantum-mechanically leads to sharp vortex-Landau levels. Using our analytic
results for the frequency-dependent complex effective mass we
conclude that in the continuum limit vortex-Landau level physics can
occur only when the system size is exponentially large compared to
the vortex core size.  In the case of many
vortices, the vortex separation would also need to be exponentially large
compared to the
average distance between vortices.
For these reasons it seems impossible
to achieve sharp vortex-Landau levels in an atom cloud
unless interaction and kinetic energies are manipulated by
an optical lattice potential.

In order to explore vortex quantum mechanics in an optical lattice
system, and as a test of our analytic continuum model results, we
have repeated our calculation for an optical lattice model in
which we are able to integrate out condensate fluctuations
numerically. Because of practical numerical restrictions we are
able to perform calculations only for system with a finite number
($\sim 2000$) of lattice sites.  Since the continuum model is
expected to apply when the vortex core size is much larger than a
lattice constant, we are not fully able to simulate the regime of
large isolated vortices for which we are able to obtain
approximate results analytically. Nevertheless, for moderately
weak interactions there is approximate agreement between our
numerical results and analytic estimates based on continuum model
calculations. As interactions strengthen we find that numerical
results for the adiabatic vortex mass continue to compare
favorably with our analytical results even when the vortex core is
smaller than an optical lattice unit cell. The discrepancy between
continuum model estimates and numerical results is much larger for
the Berry-phase coupling between orthogonal vortex displacements,
{\em i.e.} for the effective magnetic field responsible for vortex-
Landau levels. Analytic estimates of this effective magnetic field
are in fact not easily recovered numerically. This is because,
depending on the vortex core size, either the finite size of the
simulation cell cannot be ignored (for large vortex cores), or
lattice pinning is important (for small vortices).  In the strong
interaction limit we find that the effective magnetic field is
substantially reduced, lowering the Landau level frequency and
weakening its dissipative coupling to environmental modes of the
condensate.  According to our calculations, strong interactions
enhance the opportunity to observe well defined cyclotron motion
for this unexpected reason. Of course, the collective degree of
freedom in this limit is not a simple vortex displacement because
of lattice pinning effects. In addition, because the atom density
modulation associated with vortex motion is weaker at strong
interactions, it may be more difficult to couple to this
collective motion.

Although our approach treats the environmental fluctuations within
a Gaussian approximation, which is strictly speaking only valid
for weakly-interacting systems, we believe that
our conclusions are more general on a qualitative level. This is because
many of the results we find are associated with
the gapless Bogoliubov dispersion
of the environmental modes. This gaplessness is a result of
symmetry and is preserved for strong interactions.

Finally, we remark that our approach is in principle general and
can also be used to study the dissipative dynamics of other types
of topological excitations, for example, skyrmions and monopoles in
spinor Bose condensates.

\section{Acknowledgements}
This work was supported by the National Science Foundation under
the grant DMR-0606489 and by the Welch Foundation. Joseph Wang
would like to thank Jairo Sinova and Q. Niu for helpful
discussions at the early stages of this project. He also thanks
Yi-Hsiang Yu for sharing his knowledge of finite-difference methods.

\end{document}